\documentclass{article}
\usepackage{graphicx}
\usepackage{amssymb}
\title{Improving the improved action}
\date{\today}

\author{E.T. Tomboulis\footnote{e-mail: tombouli@physics.ucla.edu}  \ and
A. Velytsky\footnote{e-mail: vel@physics.ucla.edu}\\
{\em\small
Department of Physics and Astronomy, UCLA, Los Angeles, CA 90095-1547, USA}}

\begin{document}
\maketitle
\begin{abstract}
We investigate the construction of improved actions by the Monte Carlo 
Renormalization Group method in the context of $SU(2)$ gauge theory 
utilizing different decimation procedures and effective actions. 
We demonstrate that the basic self-consistency requirement for correct 
application of MCRG, i.e.  
that the decimated configurations are equilibrium configurations 
of the adopted form of the effective action, can only be achieved by 
careful fine-tuning of the choice of decimation prescription and/or 
action. 
\end{abstract}

As it is well-known, the lattice formulation is the only known 
non-perturbative formulation of gauge theories that gives the path integral 
in closed form preserving  gauge invariance and positivity of the 
transfer matrix (unitarity).  
In fact, strictly speaking, the only known way of actually defining 
`continuum' gauge theory non-perturbatively is by placing the lattice 
theory on the critical surface of a fixed point of infinite 
correlation length. Ideally, therefore, one would like to have the 
action along the Wilsonian `renormalized trajectory' which emanates 
from the fixed point and proceeds off the critical surface in only the 
relevant directions. Evolved under successive 
renormalization group (RG) 
transformations (`block-spinnings'), this `perfect action' could therefore 
be used to compute directly at any scale (coarser lattices) without any 
contamination from irrelevant directions, and hence any regularization 
artifacts, and correspondingly greatly reduced 
computational effort. Concrete practical implementation of 
this dream, however, turns out to be  rather difficult 
\cite{Hasenfratz:1997ft}.  

A more modest approach is based on the fact 
that the RG trajectory starting at any suitable lattice action will 
evolve asymptotically to the renormalized trajectory. 
Thus, after successive block spinnings one should in principle arrive at 
an `improved action' which allows computation on coarser lattices  
with suppressed discretization errors. A way to implement this is 
through the use of the Monte Carlo RG (MCRG) in which one 
performs block-spinning (decimation) transformations on gauge field 
configurations obtained by Monte Carlo simulations. 
The basic postulate here is that 
the decimated configurations are distributed according to the 
Boltzmann weight of an effective action that resulted from 
blocking to the coarser lattice. One, however, does not know a priori 
what this action is. Given a particular block-spinning prescription, the 
general procedure that has been followed is to assume a form of the effective 
action restricted to some subspace of possible interactions \cite{dF}, 
and 
then measure the couplings in this action on the decimated configurations 
by one of the known methods, the demon \cite{Creutz:1983ra} or the 
`Schwinger-Dyson' method \cite{Gonzalez-Arroyo:1986ck}.  

The purpose of this letter is to point out that, given a choice of 
decimation prescription and a choice of an effective action, 
straightforward measurement of couplings on the decimated configurations 
will, in general, lead to erroneous results. This is because the 
decimated configurations will not be equilibrium configurations 
of the effective action at these couplings. Surprisingly, this 
basic requirement underlying MCRG appears not to have been enforced in 
its application to gauge theories. Careful fine-tuning of the 
decimation prescription and/or the effective action is 
required to satisfy this requirement, if it can be satisfied at all    
within the chosen class of decimation procedures and effective actions.  

We demonstrate the presence of the problem and how it can be resolved 
in $SU(2)$ gauge theory by exploring two different decimation prescriptions: 
the Swendsen (S) 
decimation \cite{Swendsen:1981rb}, and the `Double Smeared Blocking' (DSB) 
decimation \cite{DeGrand:1994zr}. Both prescriptions involve a 
free parameter, $c$, which is the 
weight of staples relative to straight paths in the construction 
of the decimated lattice bond variable out of the undecimated lattice ones. 
We also explore two different effective action models that have been 
proposed in the literature: the multiple-representation single 
plaquette action \cite{Hasenbusch:2004yq}, \cite{Tomboulis:2005zr}, 
\cite{Tomboulis:2007rn}
\begin{equation}
S_1=\sum_{plaq}\sum_{j=1/2}^{j_N}\beta_j[1-\frac1{d_j}\chi_j(U_p)], 
\label{eq:ef_act1}
\end{equation}
and the action \cite{deForcrand:1999bi}
\begin{equation}
S_2=\sum_{plaq}\beta_{11}[1-\frac12\chi_{1/2}(U_p)]+
\sum_{rect}\beta_{12}[1-\frac12\chi_{1/2}(U_{1\times2})],
\label{eq:ef_act2}
\end{equation}
containing the single plaquette ($p=1\times1$ loop) and the $1\times 2$ 
planar loop in the fundamental representation. Here $\chi_j$ denotes the 
character of the spin-$j$ representation. 
We use the demon method to measure couplings. 

\begin{table}[ht]
 \centering
 \begin{tabular}{|c|c|c|}
  \hline
  in   &$\beta_{1/2}=\beta_{11}=$2.2578& $\beta_1=\beta_{12}=$-0.2201\\\hline
  demon &$\beta_{1/2}=$2.2582(4)&$\beta_1=$-0.2203(3)\\
  demon &$\beta_{11}=$2.2576(3)&$\beta_{12}=$-0.2201(1)\\
  \hline
 \end{tabular}
 \caption{\label{tab:dem_test}Measurements of couplings by the 
demon method before decimation.}
\end{table} 
To check the ability of the demon to measure couplings correctly, 
ensembles of 20000 configurations for the  
action (\ref{eq:ef_act1}), with $j_N=1$, and 
(\ref{eq:ef_act2}) with couplings listed in the first row 
of Table \ref{tab:dem_test} were generated on a $8^4$ lattice. 
The demon is allowed 1 sweep to set the initial energy, then 10 sweeps 
for each configuration for measurements. The results, shown in 
rows 2 and 3 of Table \ref{tab:dem_test}, demonstrate that the couplings are 
indeed accurately reproduced, though at a greater computational cost for 
(\ref{eq:ef_act2}).    

\begin{figure}[ht]
  \includegraphics[width=0.99\columnwidth]{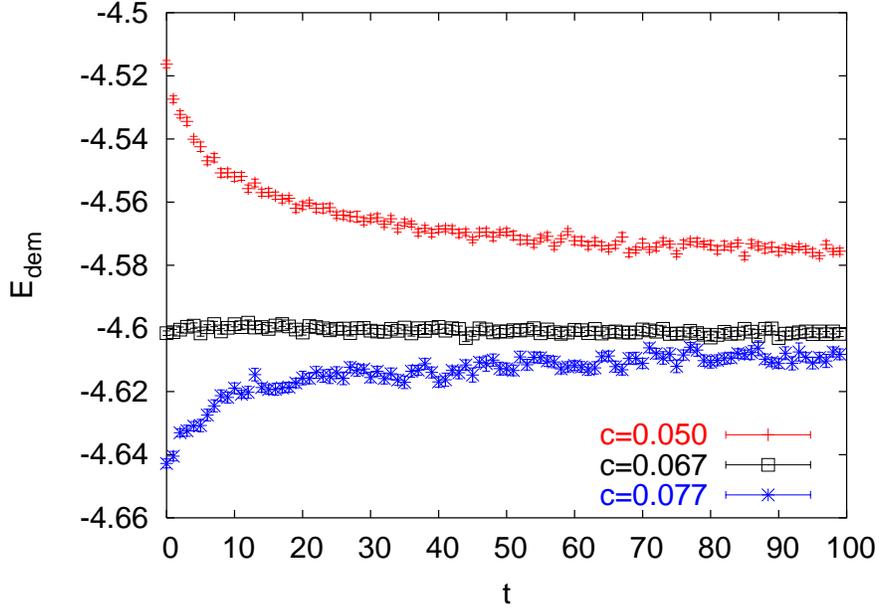}
  \caption{\label{fig:ds_dec_dem}Demon fundamental representation 
energy flow for DSB decimation and action (\ref{eq:ef_act1}) at various 
$c$ values.}
\end{figure}
Starting now with the Wilson action at coupling $\beta=2.5$ on a 
$32^4$ lattice, we perform decimations with scale factor of $2$. 
Then, adopting one of the 
effective actions (\ref{eq:ef_act1}) or (\ref{eq:ef_act2}), we proceed 
to measure the effective action couplings on the decimated configurations by 
unleashing the demon. We first consider action (\ref{eq:ef_act1}) with 
$j_N=5/2$. Fig. \ref{fig:ds_dec_dem} shows the demon fundamental 
representation energy as a function of sweeps for three values of $c$ for a 
DSB decimation. The prevalent feature 
of this plot is that there is significant energy flow during microcanonical 
evolution for two of the $c$ values shown. There is flow stabilization 
(equilibration) after about $100$ sweeps. This is in fact the 
behavior observed for any general $c$ value.  
Furthermore, this general flow pattern is typical for other representations  
\cite{Tomboulis:2007rn}. 

The implication of this is clear. Suppose one, working at some chosen 
$c$ value, measures the couplings for the effective model from 
the decimated configurations after one or a few demon sweeps (i.e. 
on the configurations as obtained right after the decimation), and 
proceeds to generate thermalized configurations of the 
effective action at {\it these} couplings. The decimated configurations 
will then {\it not} be representative of these effective action equilibrium 
configurations. As seen in Fig. \ref{fig:ds_dec_dem}
the decimated configurations will evolve under 
microcanonical evolution towards equilibration at a set of different 
values for the couplings of the effective action. But by then these 
evolved configurations no longer 
are the true original decimated configurations obtained from the underlying 
finer lattice, and cannot be generally expected to reliably preserve the 
information encoded in the original configurations. 

Ideally, one would like to have for the measurement of couplings on the 
decimated configurations the same situation as that seen in the 
test measurement of couplings on the undecimated configurations (cf. 
Table \ref{tab:dem_test} above), i.e. very fast demon thermalization 
indicating that the configurations are equilibrium configurations of 
the action for which the couplings are being measured.  
The only way out then is to seek a $c$, if any, for which this is realized. 
In the present case there is one such value, $c\in [0.065,0.067]$, as 
seen in Fig. \ref{fig:ds_dec_dem}. Furthermore, this value 
shows no significant flow, indicating very rapid thermalization, also 
for the other, in particular, the adjoint representation 
\cite{Tomboulis:2007rn}. 

A very similar state of affairs is obtained when S decimations are used in 
conjunction with (\ref{eq:ef_act1}), but the resulting fine-tuning through 
selection of a $c$-parameter value is not as sharp \cite{Tomboulis:2007rnA}.
Overall, then, DSB is better suited for the action (\ref{eq:ef_act1}), and 
can be fine-tuned so that decimated configurations are representative of 
equilibrium configurations of this action.

\begin{figure}[ht]
  \includegraphics[width=0.99\columnwidth]{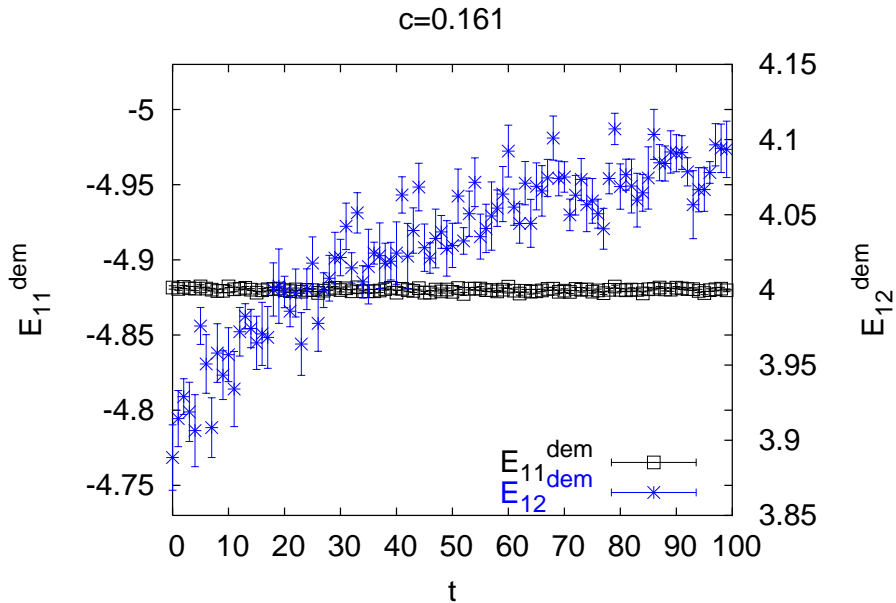}
  \caption{\label{fig:dem_ds2}Demon $1\times1$ and $1\times2$ loop
energy flow for DSB decimation at $c=0.161$.}
\end{figure}
\begin{figure}[ht]
  \includegraphics[width=0.99\columnwidth]{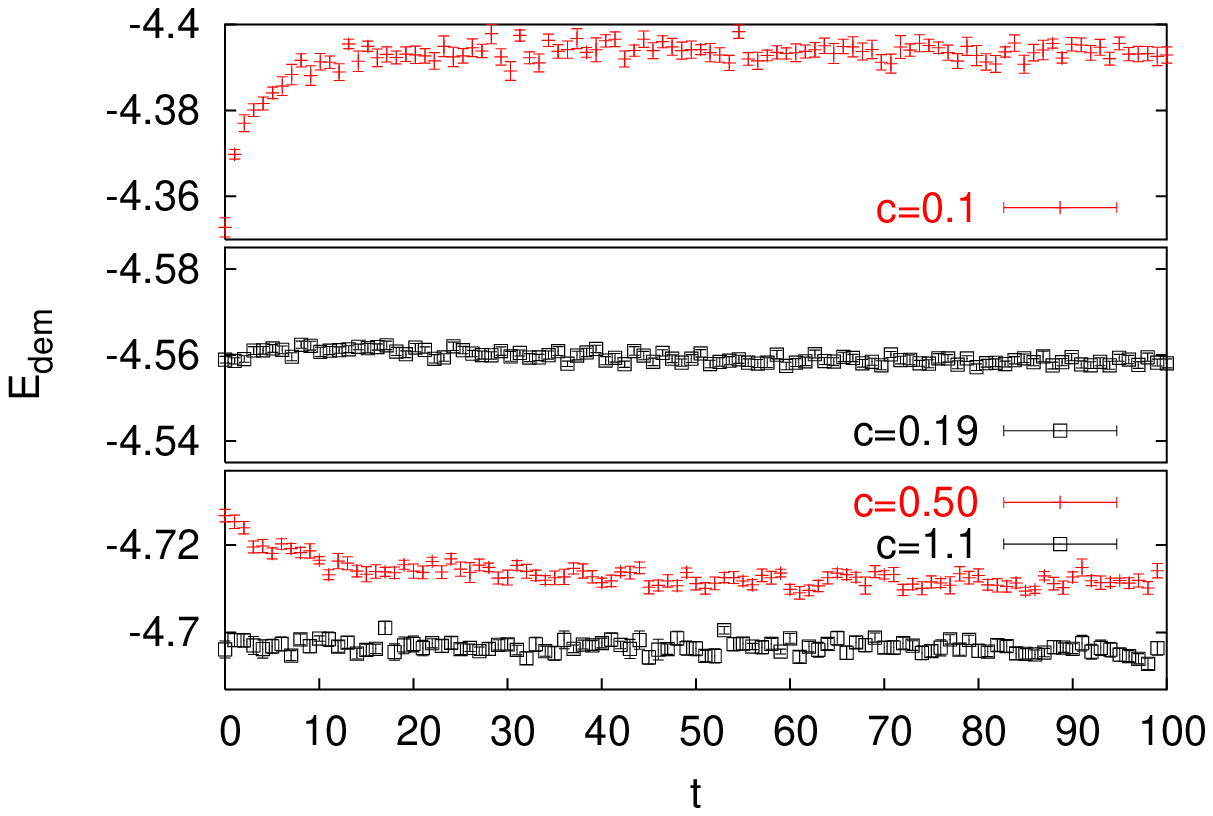}
  \caption{\label{fig:dem_sw2_11}Demon $1\times1$ loop
energy flow for Swendsen decimation at various $c$ values.}
\end{figure}
We now turn to the action (\ref{eq:ef_act2}). Employing DSB decimation, 
one again finds the same general picture for general values of the 
parameter $c$, i.e. significant flow of both 
the demon plaquette energy $E_{11}$ and $1\times2$ loop energy $E_{12}$.
There are two special $c$ values,  
$c=0.063(2)$ and $c=0.161(1)$, around which there is nearly no $E_{11}$ energy 
flow, the second one being much more sharply defined. There is, however, 
significant flow for the $E_{12}$ demon energy at both values. This is  
shown for  $c=0.161(1)$ in Fig. \ref{fig:dem_ds2}. 
Similar but stronger flow for $E_{12}$ is observed for $c=0.063(2)$  
as well. The only $c$ value for which $E_{12}$ is constant is 
$c=0.150(2)$, but $E_{11}$ shows strong flow there. It appears that there is 
no $c$ value for which DSB decimations can be fine-tuned so 
that the decimated configurations are representative of equilibrium 
configurations of the effective action (\ref{eq:ef_act2}). 
\begin{figure}[ht]
  \includegraphics[width=0.99\columnwidth]{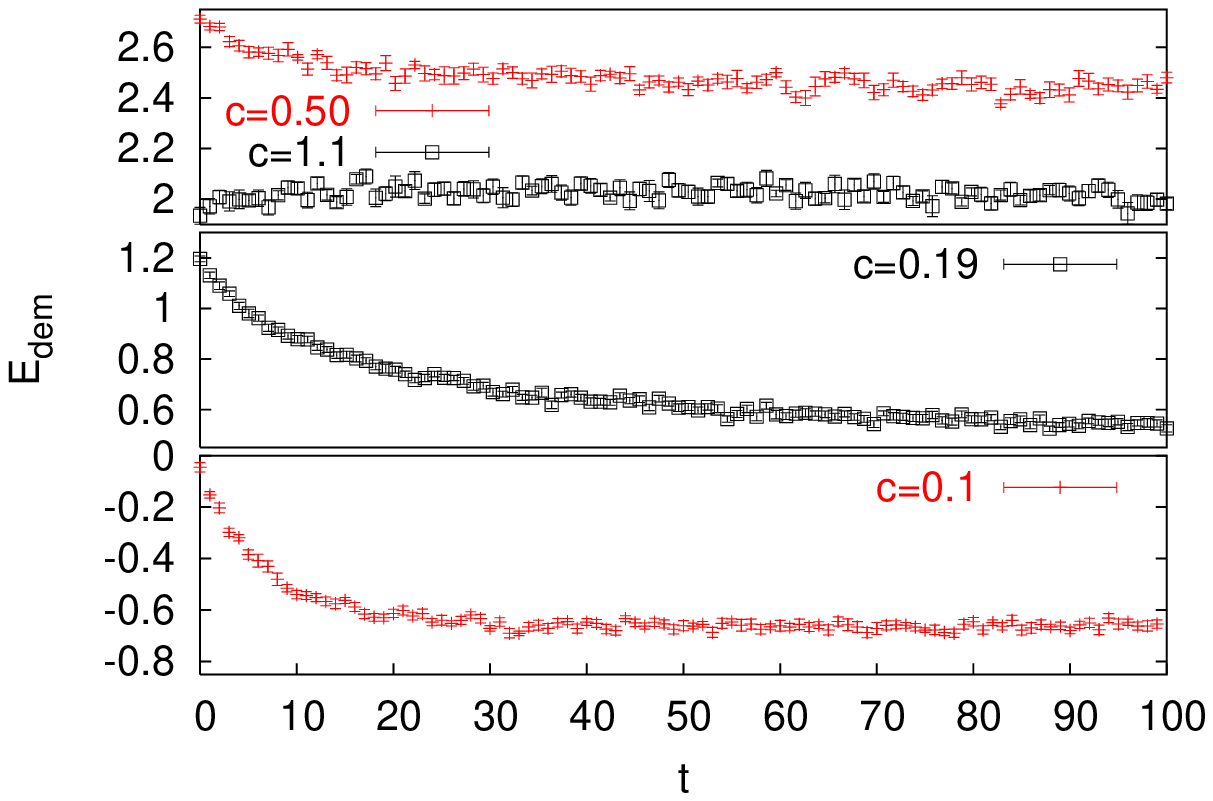}
  \caption{\label{fig:dem_sw2_12}Demon $1\times2$ loop
energy flow for Swendsen decimation at various $c$ values.}
\end{figure}
 
Next we consider Swendsen decimations with effective action 
(\ref{eq:ef_act2}). 
Figs. \ref{fig:dem_sw2_11} and \ref{fig:dem_sw2_12} present
typical evolution of demon $E_{11}$ and $E_{12}$ energies for various 
$c$ values. One sees from Fig.  \ref{fig:dem_sw2_11} that 
there is approximately no $E_{11}$ flow for $c=0.19(1)$ 
and $c=1.1(1)$. For the 
first value, however, there is a significant $E_{12}$ energy evolution as 
seen in Fig. \ref{fig:dem_sw2_12}. Thus only the latter value  
can be used. It appears then that S decimation at $c=1.1$ results in 
decimated configurations that are nearly equilibrium configurations of 
(\ref{eq:ef_act2}). 

\begin{figure}[ht]
  \includegraphics[width=0.99\columnwidth]{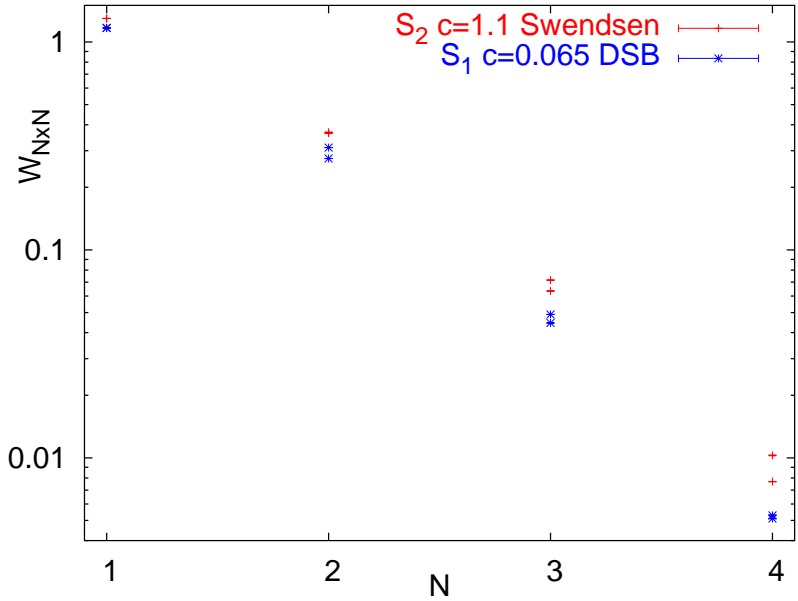}
  \caption{\label{fig:W_diff} $N\times N$ loop differences defined 
in eq. (\ref{diff}). From Table \ref{tab:W_diff}. 
}
\end{figure}

For both DSB and S decimations with effective action (\ref{eq:ef_act2}), 
the $c$-dependence of demon energies is not 
monotonic. E.g.  for DSB decimation, $E_{11}$ first it goes down, then up
with increasing $c$.   
Similar behavior is seen with S decimations (though this is not apparent in 
the selection of $c$-values shown in Fig. \ref{fig:dem_sw2_11}). Furthermore, 
the direction of the variation for  $E_{11}$ is  not consistent with that 
for $E_{12}$, as can be seen from Figs. \ref{fig:dem_sw2_11} and 
\ref{fig:dem_sw2_12}. This is in contrast  to the case of (\ref{eq:ef_act1}) 
where the demon energies vary monotonically and consistently for 
various representations and over smaller energy ranges. Furthermore,
this picture is stable under addition of higher representation terms.
This problem with action (\ref{eq:ef_act2}) may suggest that it is in fact 
unstable 
under additions of other terms in the same class of interactions, e.g.  
other loops of length size (up to) $2$ and/or other representations. 
\begin{table*}[ht]
 \centering\small
 \begin{tabular}{|c||c|c|c|c|c|}
 \hline
 $c$ &{\rm couplings}
& $R_{1\times1}$ & $R_{2\times2}$ & $R_{3\times3}$ & $R_{4\times4}$\\
 \hline \hline
 0.065&2.5023(7),-0.3098(12)&&&&\\
      &0.1057(16), -0.0397(16)&-0.0030(1)&0.1305(9)&0.106(3)&-0.034(14)\\
      &0.0145(14),-0.0029(15)&&&&\\\hline
\hline
 1.1&3.2925(6),-0.2703(2)&-0.0013(1)&-0.0096(7)&-0.115(2)&-0.252(7)\\
\hline
 \end{tabular}
 \caption{\label{tab:W_diff} Demon-measured couplings and difference of 
various size Wilson loops measured on decimated
versus effective-action-generated configurations. First row: action 
(\ref{eq:ef_act1}) with $j_N=3$ (six representations); 
second row: action (\ref{eq:ef_act2}). }
\end{table*}
To investigate this,    
and, more generally, the efficacy of (\ref{eq:ef_act1}), (\ref{eq:ef_act2}) 
as medium to long range effective actions, we consider 
some observables. In particular, for each of the actions 
(\ref{eq:ef_act1}), (\ref{eq:ef_act2}), we compare $N\times N$ loops 
measured in two ways: (a) on the decimated configurations right 
after the decimations, and denoted $W_{N\times N}^{dec}$; (b) on 
configurations generated with the effective action at couplings measured after 
decimation at the optimal $c$-value (DSB for (\ref{eq:ef_act1}),  
S for (\ref{eq:ef_act2})), and denoted $W_{N\times N}^{gen}$. 
The results for the difference  
\begin{equation}
R_{N\times N}={\Delta W_{N\times N} \over W_{N\times N}^{dec}} = {W_{N\times N}^{gen} - 
W_{N\times N}^{dec} \over W_{N\times N}^{dec}} \label{diff}
\end{equation} 
are displayed in Table \ref{tab:W_diff} and plotted in Fig. \ref{fig:W_diff}. 
In contrast to action (\ref{eq:ef_act1}), action (\ref{eq:ef_act2}) shows
consistent growth of the difference $R_{N\times N}$ with increasing $N$.
It is apparent from these results that
action (\ref{eq:ef_act2}) is 
failing as an accurate intermediate to long scale effective action, and 
must presumably be augmented by additional terms.

In conclusion, the basic self-consistency requirement for correct 
application of MCRG is 
that the decimated configurations are already equilibrium configurations 
of the adopted form of the effective action at the couplings obtained 
from  the decimated configurations. Careful fine-tuning of the decimation 
prescription and/or the effective action form is required in 
order to achieve this. 
More elaborate decimation prescriptions than the simple DSB and S 
may be defined, which involve more 
than one  adjustable parameter, and provide more fine-tuning control. 
Furthermore, more involved actions, e.g. combination of (\ref{eq:ef_act1}) and
(\ref{eq:ef_act2}), may be considered.
In particular, this program is still to be carried out for the construction 
of a reliable truly improved $SU(3)$ effective action over a wide range 
regime.

\section*{Acknowledgments}
We thank Academical Technology Services (UCLA) 
for computer support. This work was in part supported by 
NSF-PHY-0309362 and NSF-PHY-0555693.


\end{document}